\documentclass[10pt,conference]{IEEEtran}

\usepackage{amsmath} %  ,amsthm}
\usepackage{amsfonts}
\usepackage{amssymb}
\usepackage{epsfig}

\newtheorem{theorem}{Theorem}%[section]
\newtheorem{lemma}[theorem]{Lemma}

\newtheorem{observation}[theorem]{Observation}%[section]
\newtheorem{example}[theorem]{Example}%[section]
\newtheorem{problmAAA}{Problem}[section]
\newenvironment{problm}[1][\thesection]
   {\renewcommand{\theproblmAAA}{{#1}}
      \begin{problmAAA}\hspace{-3pt}}
   {\end{problmAAA}}

\makeatletter

\renewcommand{\baselinestretch}{0.95}
\begin{document}

\title{On the Minimum Number of Transmissions in Single-Hop Wireless Coding Networks}
\author{
\authorblockN{Salim Y. El Rouayheb, Mohammad Asad R. Chaudhry, and Alex Sprintson}
\authorblockA{Department of Electrical and Computer Engineering \\
Texas A\&M University, College Station, Texas \\
email:\{salim,masadch,spalex\}@ece.tamu.edu}
}

%Salim Y. El Rouayheb, Mohammad Asad R Chaudhry, and Alex Sprintson\\
%\small{Department of Electrical and Computer Engineering}\\
%\small{Texas A\&M University, College Station, Texas}\\
%\small{email:\{salim,masadch,spalex\}@ece.tamu.edu}}
\maketitle

\begin{abstract}

The advent of network coding presents promising opportunities in many areas of communication and networking. It has been recently shown that network coding technique can significantly increase the overall throughput of wireless  networks by taking advantage of their broadcast nature. In wireless networks, each  transmitted packet is broadcasted within a certain area and can be overheard by the neighboring nodes. When a node needs to transmit packets, it employs the \emph{opportunistic coding} approach that uses the knowledge of what the node's neighbors have heard in order to reduce the number of transmissions. With this approach, each transmitted packet is a linear combination of the original packets over a certain finite field.

In this paper, we focus on the fundamental problem of finding the optimal encoding for the broadcasted packets that minimizes the overall number of transmissions. We show that this problem is NP-complete over $GF(2)$ and establish several fundamental properties of the optimal solution. We also propose a simple heuristic solution for the problem based on graph coloring and present some empirical results for random settings.
\end{abstract}

\section{Introduction}

In recent years, there has been an enormous interest in the design and deployment of wireless networks. Such networks are indispensable for providing ubiquitous network coverage and have many applications in both civil and  military areas.

Recently, it was observed that the broadcast nature of wireless networks can be exploited in order to increase throughput and reduce energy consumption. In a wireless environment, each packet is broadcasted within a small neighborhood, which allows the neighboring nodes to overhear packets sent by their neighbors. When a node needs to transmit packets, it can employ the \emph{opportunistic coding} \cite{WPCPC06,KRHKMC06} approach that uses the knowledge of what the node's neighbors have heard in order to reduce the number of transmissions. With this approach, each transmitted packet is a linear combination of the original packets over a certain finite field.

\begin{figure}[t]
\begin{center}
\epsfig{file=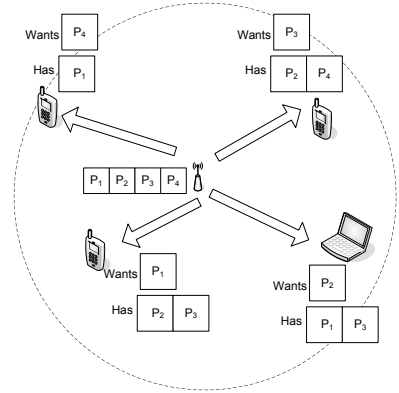, scale=0.25}
 \caption{Broadcast coding network \label{fig:op:coding}}%
\end{center}
\end{figure}

\begin{example}\label{ex:example1}
Consider the network depicted in \mbox{Figure~\ref{fig:op:coding}}. In this
example,  the central node, referred to as a server, needs to deliver four packets $p_1,\dots, p_4$ to four clients $c_1,\dots,c_4$; packet $p_i$ needs to be received by client $c_i$. Each client $c_i$  has an access to some of the packets overheard from prior transmissions. This set is referred to as its ``has'' set. It is easy to verify that all clients can be satisfied by broadcasting two packets $p_1+p_2+p_3$ and $p_1+p_4$ (all additions are over $GF(2))$. Since without network coding all packets $p_1,\dots, p_4$ are needed to be transmitted, network coding allows to reduce the number of transmissions by $50\%$.
\end{example}

% Alex: need to justify why the server knows what clients are available at the % nodes.

In this paper, we focus on the single hop wireless  setting and
consider the problem of minimizing the number of broadcast
transmissions necessary to satisfy all the clients. Our contributions
can be summarized as follows. First, we prove that the problem of determining the minimum number of transmissions over $GF(2)$ is
NP-complete. Next, we show that the number of transmissions may depend on the size of the finite field, and that such a dependence is not necessarily  monotonic. Further, we prove that the problem of finding the size of the finite field which results in the minimum number of transmissions is an NP-hard problem. Next, we establish lower and upper bounds on the \emph{coding advantage}, i.e., the ratio between the total number of packets and the minimum number of transmissions that can be achieved by using network coding.  In particular, we show that the coding advantage depends on the size of the``\emph{has}'' sets. Next, we evaluative the value of coding advantage in random settings. Finally, we present a heuristic solution based on graph coloring and verify its performance through simulations.

The considered problem is a special case of the general \emph{network
coding} \cite{ACLY00} problem for non-multicast networks. The
general network coding problem has recently attracted a large body of research (see e.g., \cite{L05,LL04} and references therein), however, many of the  results (such as NP-hardness) cannot be immediately extended to our problem.

While we present our results in the context  of wireless data
transmission, the considered problem is very general and can arise in many other practical settings. For example, consider a content distribution network that needs to deliver a set of large files (such as video clips) to different clients. In this setting, if some of the files are already available for some clients, the distribution can be efficiently implemented by multicasting a (small) set of linear combinations of the original files.

%The rest of the paper is organized as follows.  In Section
%\ref{sec:model}, we present the model and formally state the
%considered problem over finite field $GF(2)$. In Section \ref{sec:hardness}, we prove  that the problem of  finding the minimum number of transmissions over $GF(2)$ is NP-complete. In Section \ref{sec:large}, we discuss the generalization of the problem at hand for larger size finite fields.  In section
%\ref{sec:bound}, we establish lower and upper bounds on achievable
%coding gain. In Section \ref{sec:numerical}, we discuss a heuristical
%approach for minimizing the number of broadcast transmissions and
%present numerical results. Finally, the conclusions are presented in
%Section \ref{sec:conclusion}.

%\textbf{(TBC!)}

\section{Model}\label{sec:model}

We consider a one-hop wireless channel with a single server $s$ and a set of $m$ clients $C=\{c_1,\dots,c_m\}$. The server needs to transmit a set $P=\{p_1,p_2,\dots,p_n\}$ of packets to the clients. Each client requires a certain subset of packets in $P$, while some packets in $P$ are already available to it. Specifically, each client $c_i\in C$ is associated with two sets:
\begin{itemize}
\item $W(c_i)\subseteq P$ - the set of packets required by $c_i$.
\item $H(c_i)\subseteq P$ - the set of packets available at $c_i$;
\end{itemize}
We refer to $W(c_i)$ and $H(c_i)$ as the ``wants'' and ``has'' sets of $c_i$, respectively.
%
%We assume that each packet is an element of a finite field $GF(q)$. The practical issues
%
%
%
%In practical applications, large packets are divided into sub-packets of %smaller size, and the encoding is performed over the sub-packets, which allows %to reduce the encoding overhead \cite{CWJ03}.
The server can transmit any packet from $P$ as well as linear
combinations (over $GF(q)$) of packets in $P$. Each transmission $i$
is specified by an encoding vector $g_i=\{g_i^j\}\in GF(q)^n$ such that
the packet $x_i$ transmitted in communication round $i$ is equal to $x_i=\sum_{j=1}^ng_i^j\cdot p_j$.
The practical issues related to this model are discussed in \cite{KRHKMC06}.

Our goal is to find the set of encoding vectors $\Phi=\{g_i\}$ of minimum cardinality that allow each client to decode the packets it requested. We refer to this problem as Problem \emph{MIN-T-q}.

\begin{problm}[MIN-T-q]\label{prob:MIN-T}
%Let \mbox{$P=\{p_1,p_2,\dots,p_n\}$}, $p_i\in  GF(q)$ be a set of packets that need to be transmitted to a set $C=\{c_1,\dots,c_m\}$ of clients, such that the client $c_i$ needs the set $W(c_i)\subseteq P$ and has the set $H(c_i)\subseteq P$.
Find the minimum number of transmissions and the corresponding set $\Phi$ of encoding vectors  $\{g_i\}$, $g_i=\{g_i^j\}\in GF(q)^n$, that allow each client $c_i\in C$ to decode all the packets in its ``wants'' set $W(c_i)$.
\end{problm}

We assume, without loss of generality, that for each packet $p_i\in P$, there exists at least one client $c_j\in C$ such that $p_i$ belongs to the ``wants'' set $W(c_j)$ of $c_j$. We also assume that for each client $c_i\in C$ it holds that \mbox{$H(c_i)\cap W(c_i)=\emptyset$}.

%\addtocounter{theorem}{-1}

\begin{observation}
Without loss of generality, we can assume that the
``wants'' set $W(c_i)$ of each client $c_i\in C$ contains exactly
one packet. Indeed, we can substitute each client $c_i\in C$ whose ``wants'' set includes more than one packet by multiple clients $C_i=\{c_i^1,c_i^2,\dots\}$ such that the ``has'' sets of all clients in $C_i$ are equivalent to that of $c_i$ and each client in $C_i$ requests one of the packets in $W(c_i)$. It is easy to verify that the resulting instance of Problem \emph{MIN-T-q} is equivalent to the original one.
\end{observation}

\section{Hardness results}\label{sec:hardness}

In this section we focus on the case in which the encoding is performed over $GF(2)$ and prove that Problem \emph{MIN-T-2}, which is a special case of Problem \emph{MIN-T-q} for $GF(2)$, is NP-complete.

\begin{theorem}\label{theor:hard}
 Problem \emph{MIN-T-2} is NP-complete.
\end{theorem}
\begin{proof}
It is easy to verify that the problem belongs  to $NP$. To prove that
the problem is NP-complete we show a reduction from the minimum
vertex cover problem. In this problem we are given a graph $G(V,E)$
and need to find a subset $\hat{V}$ of $V$, of minimum cardinality, such
that each edge $e\in E$ is incident to at least one of the nodes in
$\hat{V}$. We denote by $OPT^{VC}=|\hat{V}|$ the size of the optimal solution
for the vertex cover problem.

Given an instance $G(V,E)$ to the vertex  cover problem we build the
following instance for Problem \mbox{\emph{MIN-T-2}}.
%We define a packet set as $P=\{p_v\ | v\in V\}\cap \{p_e\ | e\in E\}$,
The packet set $P$ includes a packet $p_v$ for any node in $V$  and
a packet $p_e$ for any edge in $E$.  We denote by $P_V=\{p_v\ |\ v\in
V\}$ the subset of packets in $P$ that correspond to nodes in $V$
and by $P_E=\{p_e\ |\ e\in E\}$ the subset of packets in $P$ that
correspond to edges in $E$.

%Let $n=|X_E\cup X_V|$.

For each edge $e(v,u)\in E$ we define two clients $c_e^1$ and $c_e^2$ such that:
\begin{itemize}
\item $H(c_e^1)=\{p_e\}$ and $W(c_e^1)=\{p_v,p_u\}$;
%\item $H(c_e^2)=\{e,v\}$, $W(c_e^1)=\{u\}$;
%\item $H(c_e^3)=\{e,u\}$, $W(c_e^1)=\{v\}$;
\item $H(c_e^2)=\{p_v,p_u\}$ and $W(c_e^2)=\{p_e\}$.
\end{itemize}

We denote by $OPT$ the size of the optimal solution for this instance of Problem \mbox{\emph{MIN-T-2}}, i.e., the minimum number of transmissions necessary to satisfy all clients. In the following two lemmas we prove that \mbox{$OPT=OPT^{VC}+|E|$}.
\begin{lemma}\label{lemm:1}
$OPT\leq OPT^{VC}+|E|$.
\end{lemma}
\begin{proof}
Let $\hat{V}\subseteq V$ be the optimal solution to  the vertex cover problem. Then, all clients can be satisfied by transmitting the following set of packets of size $OPT^{VC}+|E|$:
\begin{enumerate}
\item For each node $v\in \hat{V}$  we transmit the corresponding packet $p_v$;

\item For each edge $e(v,u)\in E$ we transmit the packet  $p_v+p_u+p_e$, where $p_v$, $p_u$, and $p_e$ are packets that correspond to nodes $v$, $u$, and edge $e$, respectively.
\end{enumerate}

It is easy to verify that the set $\Phi$ of corresponding encoding vectors is a feasible solution to Problem \mbox{\emph{MIN-T-2}}.  Since the total number of transmitted packets is $OPT^{VC}+|E|$ it follows that  $OPT\leq OPT^{VC}+|E|$.
\end{proof}

%\newpage

\begin{lemma}\label{lemm:2}
$OPT\geq OPT^{VC}+|E|$.
\end{lemma}
\begin{proof}
Consider an optimal solution $\Phi=\{g_1,\dots, g_{OPT}\}$ to Problem~\mbox{\emph{MIN-T-2}}, where $$g_i=(g_i^{v_1}, \dots,g_i^{v_{|V|}},
g_i^{e_1},\dots,g_i^{e_{|E|}})\in GF(2)^{|E|+|V|}.$$

With this solution, the packet transmitted at round $i$ is equal to

$$x_i=\sum_{v_j\in V} g_i^{v_j}\cdot p_{v_j}+\sum_{e_j\in E} g_i^{e_j}\cdot p_{e_j}.$$

We denote by
$\langle \Phi\rangle $ the linear subspace of dimension $OPT$ of
$GF(2)^{|V|+|E|}$ generated by the vectors in $\Phi$.

%
%$X=\{x_1,\dots,x_{OPT}\}$,
%$x_i=\sum_{v\in V}g_i^v\cdot p_v+\sum_{e\in E}g_i^e\cdot p_e$ to
%Problem \emph{MIN-T-2}, where

%We denote by $H=\{g_1,\dots, g_{OPT}\}$
%the set of encoding vectors of packets in $P$, where
%$$g_i=(g_i^{v_1}, \dots,g_i^{v_{|V|}},
%g_i^{e_1},\dots,g_i^{e_{|E|}})\in GF(2)^{|E|+|V|}.$$ We denote by
%$\langle H\rangle $ the linear subspace of dimension $OPT$ of
%$GF(2)^{|V|+|E|}$ generated by the vectors in $H$.

We show that there exist  two sets $\Phi_1$ and $\Phi_2$ of vectors in $\langle \Phi\rangle $ and a vertex cover $\hat{V}\subseteq V$  such that the following three conditions holds:
\begin{enumerate}
\item[(1)] For any edge $e\in E$,  there exists an encoding vector $g_i\in \Phi_1$ such that $g^e_i=1$, and $g^{e'}_i=0$ for any edge $e'\in E\setminus\{e\}$;
\item[(2)] For each $g_i\in \Phi_2$ it holds that $g^e_i=0$ for any edge $e\in E$;

\item[(3)] For each $v\in \hat{V}$ there exists an encoding vector $g_i\in \Phi_2$ such that $g^v_i=1$ and $g^{v'}_i=0$ for any node $v'\in \hat{V}\setminus \{v\}$.
\end{enumerate}

Note that all encoding vectors in $\Phi_1\cup \Phi_2$ are linearly independent, $|\Phi_1|=|E|$, and $|\Phi_2|=|\hat{V}|$.

First, we show how to construct the set $\Phi_1$. Let $e(v,u)$ be an edge in $E$ and let $c_e^1$ and $c_e^2$ be the two clients that correspond to $e$. We note that in order to satisfy $c_e^2$, $\langle \Phi\rangle $ must contain at least one vector $g_i$  for which it  holds $g_i^e=1$ and $g^{e'}_i=0$ for any
edge $e'\in E\setminus\{e\}$. Thus, we can form $\Phi_1$ by including, for each $e\in E$, the vector $g_i\in\langle \Phi\rangle $ that corresponds to
$e$.
%Clearly, $\Phi_1$ contains $|E|$ linearly independent vectors.

Second, we show how to construct set $\Phi_2$ and the vertex cover $\hat{V}$. Again, let $e(v,u)$ be an edge in $E$ and let $c_e^1$ and $c_e^2$ be the two clients that correspond to $e$. Note that, in order to satisfy the client $c_e^1$, the set $\langle \Phi\rangle $ must contain a vector $g_i$ for which it holds that $g^{e'}_i=0$ for all $e'\in E$, $g^w_i=0$ for all $w\in V\setminus \{v,u\}$, and either $g^v_i$ or $g^u_i$ (or both) are non-zero. Let
$T$ be a set that contains such vectors for all $e\in E$. Let
\mbox{$l=\dim \langle T\rangle $}. It follows from linear algebra
that there exists an  $l\times (|V|+|E|)$ matrix $M$ over $GF(2)$
that satisfies the following conditions:
\begin{enumerate}
\item The rows of $M$ span $\langle T\rangle $;
\item  There are $l$ linearly independent columns in $M$ such that each column contains exactly one non-zero element.
\end{enumerate}

Indeed, we can first construct an $l\times (|V|+|E|)$ matrix $M'$ whose rows span $T$. Such matrix is of rank $l$, hence it contains at least $l$ non-zero columns which are linearly independent. The matrix $M$ can be constructed form $M'$ by performing Gaussian elimination. We denote by $\hat{V}$ the subset of $V$ that corresponds to $l$ linearly independent columns of $M$, each column contains exactly one non-zero element. Then, we set $\Phi_2$ to be the set of row vectors of $M$. Note that $\Phi_2$ has $l$ elements.

We proceed to show that $\hat{V}$ is a vertex cover in $G(V,E)$. We note that the structure of $M$ implies that for any non-zero vector $g_i$ in the row span of $M$, and, in turn, in $\langle T\rangle $ it must hold that $g_i^w=1$ for some $w\in \hat{V}$. For each edge $e(v,u)\in E$ let $g_i$ be the vector that correspond to $e$ in $T$. Recall $g_i$ has one or two non-zero components, which are either $g_i^v$ or $g_i^u$, or both. This implies that either $v$ or $u$, or both belong to $\hat{V}$.

%Suppose, by way of contradiction,  that there exists an edge $e(v,u)\in E$ such that $v\notin \hat{V}$ and $u\notin \hat{V}$.  This, however, contradicts the fact that $T$ contains a vector $g_i$ for which it holds that $g^e_i=0$ for $e\in E$, $g^w_i=0$ for all $w\in V\setminus \{v,u\}$,  and either $g^v_i$ or $g^u_i$ (or both) are non-zero.

We proved that there exist two sets $\Phi_1$ and $\Phi_2$ of independent  vectors in $\langle \Phi\rangle $ such that $|\Phi_1|=|E|$, and $|\Phi_2|\geq OPT^{VC}$. We conclude that
 $$OPT=\dim\langle \Phi\rangle \geq|\Phi_1|+|\Phi_2|\geq OPT^{VC}+|E|.$$
\end{proof}

From lemmas \ref{lemm:1} and \ref{lemm:2} it follows that $OPT= OPT^{VC}+|E|$. Thus,  a polynomial-time algorithm that solves Problem \emph{MIN-T-2} will solve the vertex cover problem as well, resulting in a contradiction.
\end{proof}

\section{Dependence on the field size}\label{sec:large}

In this section we consider a variant of  Problem~\mbox{\emph{MIN-T-q}}
which allows flexibility in choosing the underlying finite filed. $GF(q)$. Specifically, for each instance of the problem, we can choose the finite field that minimizes the required number of transmissions.

We denote by $OPT(q)$ the minimum required number of transmissions over $GF(q)$. We also denote by $OPT$ the minimum number of transmissions that can be achieved over any finite field.

 %of packets $P$ and set
%
%
%
%In this section, we study the
%
%For  specific set of packets, clients and demands, we define
%$OPT(q)$ to be minimum number of transmissions needed to solve the
%corresponding  Problem \emph{MIN-T-q}. In addition, we denote by  $OPT^*$ the minimum number of transmission needed in general when there is no
%constraint on the field $GF(q)$; i.e.\ $$OPT^*=\min_q OPT(q).$$

We begin by observing that the minimum number of transmissions may depend on the size of the finite field $GF(q)$. For example, consider the problem described in Table \ref{tab:ex2}, where $P=\{p_1,p_2,p_3,p_4\}$, and for every client $c_i$, $H(c_i)=P\setminus W(c_i)$. We prove that in this problem $OPT(2)>OPT(3)$.

%\begin{example}\label{ex:example2}
%Consider the

First, we show that $OPT(2)>2$. Suppose, by way of contradiction, that there is a solution  to this problem with two transmissions:
\begin{equation*}
\begin{array}{c}
x_1=g^1_1p_1+\dots+g_1^4p_4 \\
x_2=g_2^1p_1+\dots+g_2^4p_4
\end{array}
\end{equation*}
To satisfy all clients, the vectors $(g^1_1,g_2^1),\dots,(g_1^4,g_2^4)$ should be all distinct and different from $(0,0)$, which is not possible over $GF(2)$. Since the set of transmissions $\{p_1+p_3, p_2+p_3, p_4\}$ satisfies all clients, it follows  that $OPT(2)=3$. We note that $OPT$ is at least two, since $OPT\geq |W(c_i)|=2$. We also observe that over $GF(3)$ only two transmissions
$\{p_1+p_3+p_4, p_2+p_3+2p_4\}$ are sufficient, hence $OPT(3)=OPT=2$.
%\end{example}

%The next lemma shows that the optimal solution depends on the field size $q$.

%\begin{lemma}
%The is an instance for the broadcast problem for which it holds that $OPT(2)<OPT(3)$.
%\end{lemma}
%\begin{proof}

%$$x_1=g^1_1p_1+\dots+g_1^4p_4$$ and $$x_2=g_2^1p_1+\dots+g_2^4p_4.$$

\begin{table}[!htb]
  \centering
  \begin{tabular}{|c|c|c|}
  \hline
 $C$ & $W(c_i)$& $H(c_i)$ \\
  \hline\hline
 $c_1$& $\{p_1,p_2\}$ & $\{p_3,p_4\}$ \\\hline
 $c_2$& $\{p_1,p_3\}$ & $\{p_2,p_4\}$ \\\hline
 $c_3$& $\{p_1,p_4\}$ & $\{p_2,p_3\}$\\\hline
  $c_4$& $\{p_2,p_3\}$ & $\{p_1,p_4\}$ \\\hline
 $c_5$& $\{p_2,p_4\}$ & $\{p_1,p_3\}$\\\hline
 $c_6$& $\{p_3,p_4\}$ & $\{p_1,p_2\}$\\\hline
 \end{tabular}
 \caption{ %Instance of a problem where $n=4$ and  $OPT(3)<OPT(2)$.
 }\label{tab:ex2}
\end{table}
%\end{proof}

The following lemma shows that $OPT(q)$ is not necessarily a monotonic function of $q$.

\begin{lemma}
There exists an instance of  Problem~\mbox{\emph{MIN-T-q}} for which it holds that $OPT(q)=3$ for fields with odd characteristic, such as $GF(3)$ and $OPT(q)>3$ for fields with even characteristic.
\end{lemma}
%\begin{proof}(sketch)
\noindent\hspace{2em}{\it Proof (sketch):}
Consider the problem instance described in Table \ref{tab:ex3}, where
$P=\{p_1,\dots,p_{7}\}$, and \mbox{$H(c_i)=P\setminus W(c_i)$}.

\begin{table}[!htb]
  \centering
  \begin{tabular}{|c|c|c|}
  \hline
 $c_i$ &  $W(c_i)$\\
  \hline\hline
 $c_1$& $\{p_1\}$  \\\hline
 $c_2$& $\{p_2\}$ \\\hline
  $c_3$& $\{p_3\}$ \\\hline
   $c_4$& $\{p_2,p_4\}$  \\\hline
 $c_5$& $\{p_3,p_5\}$ \\\hline
  $c_6$& $\{p_3,p_6\}$ \\\hline
  $c_7$& $\{p_4,p_7\}$ \\\hline
  $c_8$& $\{p_5,p_7\}$ \\\hline
  $c_9$& $\{p_6,p_7\}$ \\\hline
  $c_{10}$& $\{p_4,p_5,p_6\}$ \\\hline

 \end{tabular}
 \caption{ %Instance ($n=7$) where $OPT(q)=OPT=3$ only when $q$ is an odd prime % power
 }\label{tab:ex3}
\end{table}
For fields with odd characteristic,  the transmission sequence
$$\{p_1+p_4+p_5+p_7, p_2+p_4+p_6+p_7, p_3+p_5+p_6+p_7 \}$$ satisfies all clients, hence $OPT(q)=OPT=3$.

%
%Note that this solution does not work over fields with even
%characteristic, since the client $c_{10}$ will need to invert the
%system $\{p_4+p_5, p_4+p_6, p_5+p_6\}$ which  is not possible
%over $GF(2^k)$.

We observe that for fields with even characteristic ($q=2^k$) it holds that
$OPT(2^k)>3$. Indeed, for any solution \mbox{$\{g_1,g_2,g_3\}\in
GF(q)^3$} with three  transmissions, consider the matrix $T$ whose
row vectors are $g_1$, $g_2$, and $g_3$. To satisfy the demands of all the clients the vector matroid of $T$ should be isomorphic to the Fano matroid
\cite{OX93}. But, the Fano matroid is only representable over fields
with odd characteristics. Therefore, there are no solutions to above
problem with three transmissions over $GF(2^k)$.
\hspace*{\fill}~\QED\par\endtrivlist\unskip
%\end{proof}

The next lemma shows that deciding whether the optimal number of transmissions can be achieved for a given field $GF(q)$ is an NP-hard problem.

\begin{lemma}
For given a prime power $q$, it is an NP-hard problem to decide whether $OPT(q)=OPT$.
\end{lemma}
%\begin{proof}(sketch)
\noindent\hspace{2em}{\it Proof (sketch):} Similar to \cite{LL04},
we use a reduction from the problem of graph coloring. Given an
undirected graph $G(V,E)$, we construct the following instance to
the broadcast problem. For each node $v\in V$, the set $P$ includes
a packet $p_v$. For each edge $e(v,u)\in E$, the set $C$ includes a
client $c_e$ such that $W(c_e)=\{p_v,p_u\}$ and $H(c_e)=P\setminus
W(c_e)$. It is easy to verify that for this problem it holds that
$OPT=2$.

We show the problem can be solved with two transmissions over $GF(q)$ if and only if $G$ is $q+1$ colorable.
First, suppose that $G(V,E)$ can be colored with $q+1$ colors. Let $d(v)\in\{1,\dots,q+1\}$ be the
color of vertex $v$. As shown in \cite{LL04}, there exists $q+1$ pairwise independent vectors
$(z_1^1,z_2^1),\dots,(z_1^{q+1},z_2^{q+1})$ over $GF(q)$. For each node  $v\in V$ we set
 $(g_1^v,g_2^v)=(z_1^{d(v)},z_2^{d(v)})$. It is easy to verify that  the two encoding
 vectors $(g_1^v)_{v\in V}$ and $(g_2^v)_{v\in V}$
   constitute a feasible solution for the broadcast problem.

Second, suppose that there exists a solution $\Phi=\{(g_1^v)_{v\in
V}, (g_2^v)_{v\in V} \}$ for the broadcast problem with two
transmissions. We show that this implies that there exists a $q+1$
coloring of graph $G$. For each vertex $v\in V$, the vector
$(g_1^v,g_2^v)$ determines the coefficients for packet $p_v$ for the
first and
 the second transmissions  in $\Phi$. The set of such vectors can be partitioned into $q+1$ equivalence classes, such
 that any two linearly dependent vectors are placed into the same equivalence class. Next, for each equivalence class we
 assign one of the $q+1$ colors.  Next,  for each vertex $v\in V$ we assign the color that corresponds to the equivalence
  class of $(g_1^v,g_2^v)$. It is easy to verify that this will result in a valid coloring of $G$ that requires at most
  $q+1$ colors.
\hspace*{\fill}~\QED\par\endtrivlist\unskip
%\end{proof}

\section{Bounds on coding advantage}\label{sec:bound}

Given a one-hop transmission problem with $n$ packets $P=\{p_1,p_2,\dots,p_n\}$ and $m$ clients $C=\{c_1,\dots,c_m\}$, we define the coding gain $\Gamma$ as the ratio between the minimum number of transmissions without coding and the minimum number of transmissions with  coding, i.e.,
$$\Gamma=\frac{n}{OPT},$$
where $OPT$ is the minimum number of transmissions achievable over any finite field $GF(q)$.

Let $L=\max_{{c_i}\in C} |H(c_i)|$ and $\ell=\min_{c_i\in C}|H(c_i)|$. The following theorem establishes lower and upper bounds on $\Gamma$.

\begin{theorem}
The coding gain is bounded by
\begin{equation}\label{bounds}
   \frac{n}{n-\ell}\leq \Gamma \leq L+1
\end{equation}

\begin{proof}
We assume, without loss of generality, that $|W(c_i)|=1$
for each  client $c_i\in C$. Let $GF(q)$ be the field that requires $OPT$ transmissions. Consider an optimum solution $\Phi$ that includes $OPT$ encoding vectors $g_1,\dots,g_{OPT}$.

Let $e_j\in GF(q)^n$, $1\leq j\leq n$, be the unit vector whose
components are all zeros except for the $j$-th one which is 1. For
$1\leq i\leq n$, we define $w_i= e_j$ if client $c_i$ wants packet
$p_j$. Also, we define
$$\hat{H}(c_i)=\{e_j; p_j\in H(c_i)\}.$$

To guaranty that each client $c_i$ is able to decode the
packet $h_i$ in its ``wants'' list, there must be a vector $y_i\in \langle \Phi\rangle $ such that $w_i=y_i+h_i$,  $h_i\in \langle\hat{H}(c_i)\rangle$, where $\langle \Phi\rangle $ and $\langle\hat{H}(c_i)\rangle$ are the  linear subspaces generated by the vectors in $\Phi$ and $\hat{H}(c_i)$, respectively. We note that the Hamming weight of $y_i$ is upper bounded by $L+1$.

Let $Y=\{y_i\ | \ c_i\in C \}$. By the optimality of the
solution the dimension of the linear subspace $\langle Y\rangle$ generated by $Y$ is equal to that of $\langle \Phi\rangle $. Let $B$ be the $OPT \times n$ matrix whose row vectors belong to $Y$ and form the basis of $Y$.
 We note that $B$ must satisfy the following two conditions:
\begin{enumerate}
  \item Each row of $B$ contains at most $L+1$ non-zero elements.
  \item $B$ does not contain the all-zero column vector.
\end{enumerate}
The first condition follows from the upper bound on the Hamming
weights of the vectors in $Y$. The second condition follows from the
observation that for every packet $p_i$ at the source, there is at
least one client that wants it. These two conditions imply that $OPT \geq \frac{n}{L+1}$.

We proceed with the lower bound. Given an instance $I_1$ of Problem \emph{MIN-T-q}, we form another instance $I_2$
with where all the ``has'' sets have order $\ell$, and where
$W(c_i)=P\setminus H(c_i)$. Instance $I_2$ is formed by deleting arbitrary elements from the ``has'' sets of $I_1$ and expanding the ``wants'' sets of its elements. Note that any valid solution for instance $I_2$ is also a valid solution for instance $I_1$.

For a field $GF(q)$ of large enough size (larger than the number of
clients), we can always find a subspace $S$ of dimension $n-\ell$ in
$GF(q)^n$ that is simultaneously orthogonal to all the subspaces
$<\hat{H} (c_i)>$ corresponding to  $I_2$ (Theorem 1 in \cite{CAYE01}).
Any basis of $S$ will constitute a solution for $I_2$, and, in turn, for
$I_1$, which requires $n-\ell$ transmissions. Thus, the lower bound follows.
\end{proof}

\end{theorem}

\section{Heuristic Approach and Numerical Results}\label{sec:numerical}

\subsection{Heuristic Approach}\label{sec:heuristic}

In Section \ref{sec:hardness}, we proved that Problem~\emph{MIN-T-2} is NP-complete, hence finding an optimal solution for large instances of the problem can be impractical. In this section, we present a heuristic approach to solve  this problem. Our heuristic solution employs \emph{memoryless decoding}, i.e., each client uses exactly one of the transmitted packets to decode one of the packets in its ``wants'' list and never uses a linear combination of the transmitted packets. While memoryless decoding, in general, results in a suboptimal solution, our numerical results, presented below, show that in many cases the number of required packets is close to the optimum.  We observe that the problem of finding the minimum number of transmissions with memoryless decoding is equivalent to the problem of finding the minimum chromatic number of an undirected graph.

Specifically, consider an instance $I$ problem of Problem
\emph{MIN-T-q}, in which the ``wants'' set of each client is of
cardinality one. Then, we construct an instance $G(V,E)$ to graph
coloring problem through the following procedure:

\begin{itemize}
\item For each client $c_i\in C$ there is a corresponding vertex $v_{c_i}$  in $V$
\item Each two vertices $v_{c_i}$ and $v_{c_j}$ are connected by an edge if one of the following holds:
    \begin{itemize}
        \item Clients $c_i$ and $c_j$ have identical ``wants'' sets;

        \item $W(c_i)\subseteq H(c_j)$ and $W(c_j)\subseteq H(c_i)$.

    \end{itemize}
\end{itemize}

Let $\hat{V}\subseteq V$ be a clique in $G(V,E)$, i.e., each two vertices
of $V$ are connected by an edge in $G$. Note that all clients that
correspond to nodes in $\hat{V}$ can be satisfied by one transmission,
which includes a linear combination of all packets in their
``wants'' sets. Thus, the minimum number of transmissions with
memoryless decoding can be found by solving a \emph{clique partition}
problem \cite{GJ79}, i.e., partition of $V$ into disjoint subsets
$V_1,V_2,\dots, V_k$, such that for $1\leq i\leq k$, the subgraph of
$G$ induced by $V_i$ is a complete graph. This problem, in turn,
corresponds to the minimum graph coloring problem of the
complimentary graph. The latter problem is a well-studied problem
with a wealth of heuristic solutions developed in the recent years.

%Figure \ref{fig:op:coding7} depicts the values of the coding gain that can be achieved through heuristic approach.
%
%
%\begin{figure}[t]
%\begin{center}
%\epsfig{file=fig7.eps, scale=0.35}
% \caption{Average coding gain  using greedy heuristic\label{fig:op:coding7}}%
%\end{center}
%\end{figure}

\begin{figure}[t]
\begin{center}
\epsfig{file=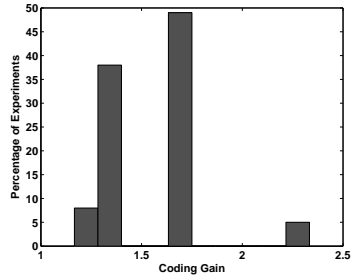, scale=0.35}
 \caption{Histogram of coding gain for 7 clients with optimal decoding \label{fig:op:coding2}}%
\end{center}
\end{figure}

\begin{figure}[t]
\begin{center}
\epsfig{file=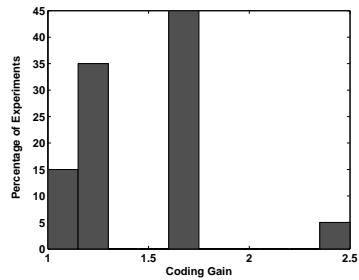, scale=0.35}
 \caption{Histogram of coding gain for 5 clients using memoryless decoding \label{fig:op:coding4}}%
\end{center}
\end{figure}

\subsection{Numerical results}

We performed several numerical experiments  in order to evaluate the
coding gain as well as the performance of the heuristic solution in
random settings. In all of our experiments described below, the
``wants'' set of each client is of cardinality one, and the number
of clients is equal to the number of packets.

In the first experiment, we evaluated the coding  gain of a
single-hop wireless system with seven clients. Specifically, we
generated 50 instances of Problem \emph{MIN-T-q}, in each setting
the set ``has'' of each client is randomly selected. The results of
the experiment are shown in Figure \ref{fig:op:coding2}. The results
show that in the majority of the experiments, there is a significant
coding gain (more than 1.75).

The second experiment is similar to  the first one, but the clients
only employ memoryless decoding.  The results of the experiment are
shown in Figure \ref{fig:op:coding4}. The results show that a
significant coding gain (up to 2.5) can be achieved, while in the
majority of the cases, the coding gain is at least 1.7.

\begin{figure}[t]
\begin{center}
\epsfig{file=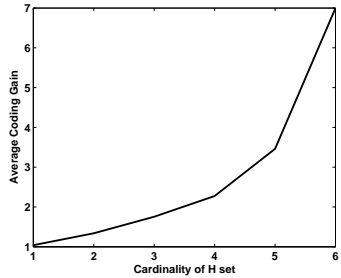, scale=0.35}
 \caption{Average Coding gain as a function of the cardinality of the ``has'' set.\label{fig:op:coding3}}%
\end{center}
\end{figure}

\begin{figure}[t]
\begin{center}
\epsfig{file=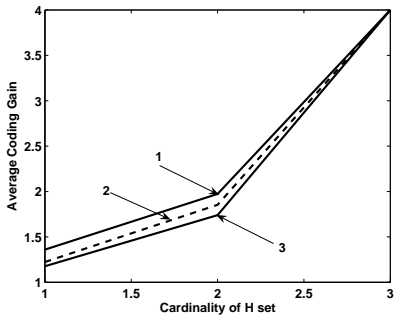, scale=0.4}
 \caption{Average Coding gain as a function of the cardinality of the ``has'' set using different techniques: (1) optimal decoding; (2) memoryless decoding; (3) heuristic approach.\label{fig:op:coding5}}%
\end{center}
\end{figure}

In the third experiment, we studied  the dependence of average
coding gain on the cardinality of the ``has'' sets. In particular,
we generated a problem instance in which the cardinality of the
``has'' set is equal for all clients, while the content of the
``has'' set was randomly selected. Figure \ref{fig:op:coding3} shows
the average coding gains of the system with seven clients using
optimal decoding as a function of cardinality of the "has" sets,
while Figure \ref{fig:op:coding5} shows the comparison of average
coding gains as a function of cardinality of the "has" sets, for
three techniques i.e., optimal decoding, memoryless decoding and
heuristic approach.The results show that the average coding gain
increases with the size of the ``has'' sets, which confirms the
intuition that coding is more beneficial if the clients have more
packets in their ``has'' sets.

\begin{figure}[t]
\begin{center}
\epsfig{file=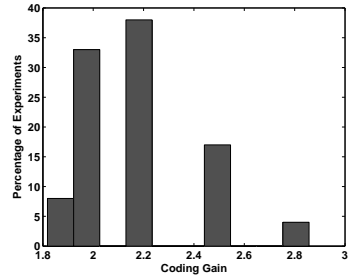, scale=0.35}
 \caption{Histogram of coding gain for 20 clients using heuristic approach\label{fig:op:coding7}}%
\end{center}
\end{figure}

Finally, we evaluated the coding gain  that can be obtained through
the heuristic approach presented in Section \ref{sec:heuristic}. The
results of this experiment are depicted in Figure
\ref{fig:op:coding7}. The results show that the proposed heuristic
approach allows to obtain a significant reduction in the number of
transmitted packets.

%\section{Heuristic Approach}

%\begin{figure}[t]
%\begin{center}
%\epsfig{file=fig6.eps, scale=0.35}
% \caption{Avg. Coding gain-vs- Cardinality of H  for 4 clients using greedy heuristic\label{fig:op:coding}}%
%\end{center}
%\end{figure}

\section{Conclusion}\label{sec:conclusion}

The paper focuses on minimizing the  number of transmissions
necessary for satisfying all clients in single-hop wireless settings.
We employ the technique of network coding which allows to
take advantage of the packets that were overheard from  prior
transmissions.

Our paper makes the following contributions.  First, we proved that
the problem of finding the minimum number of transmissions is
NP-complete over the binary field. Second, we analyzed an extended
version of the problem in which the encoding can be performed over a
larger finite field. Furthermore, we established lower and upper
bounds on the value of the  coding gain. Next, we presented a
heuristic solution based on graph coloring. Finally, we conducted a simulation study that evaluates the coding gains in practical settings.

The considered problem presents significant challenges and provides a fertile ground for future research. In particular, we would like to prove the NP-hardness and inapproximability for finite fields of larger size as well as for non-linear network codes.

\bibliographystyle{unsrt}
\bibliography{article}

\end{document}